\documentclass[11pt,a4paper]{article}
\pdfoutput=1
\usepackage{jheppub}
\newcommand{\Ein}{E_\mathrm{in}}
\newcommand{\Eout}{E_\mathrm{out}}
\newcommand{\be}{\begin{equation}}
\newcommand{\ee}{\end{equation}}
\newcommand{\ba}{\begin{eqnarray}}
\newcommand{\ea}{\end{eqnarray}}
\renewcommand{\(}{\left(}
\renewcommand{\)}{\right)}

\newcommand{\lk}{\left[}
\newcommand{\rk}{\right]}

\newcommand{\w}{\omega}
\newcommand{\tf}{\tilde{f}}
\newcommand{\hw}{\hat{\omega}}
\newcommand{\nn}{\nonumber}

\title{Holographic dilepton production in a thermalizing plasma}

\author[a]{Rudolf Baier,}
\author[b]{Stefan A. Stricker,}
\author[a]{Olli Taanila,}
\author[a]{and Aleksi Vuorinen}

\affiliation[a]{Faculty of Physics, University of Bielefeld,\\
D-33615 Bielefeld, Germany}
\affiliation[b]{Institute of Theoretical Physics, Technical University of Vienna,\\
Wiedner Hauptstr.~8-10, A-1040 Vienna, Austria}

\emailAdd{baier@physik.uni-bielefeld.de}
\emailAdd{stricker@hep.itp.tuwien.ac.at}
\emailAdd{olli.taanila@iki.fi}
\emailAdd{vuorinen@physik.uni-bielefeld.de}

\abstract{We use the AdS/CFT correspondence to determine the out-of-equilibrium production rate of dileptons at rest in strongly coupled ${\mathcal N}=4$ Super Yang-Mills plasma. Thermalization is achieved via the gravitational collapse of a thin shell of matter in AdS$_5$ space and the subsequent formation of a black hole, which we describe in a quasistatic approximation. Prior to thermalization, the dilepton spectral function is observed to oscillate as a function of frequency, but the amplitude of the oscillations decreases when thermal equilibrium is approached. At the same time, we follow the flow of the quasinormal spectrum of the corresponding U(1) vector field towards its equilibrium limit.}


\begin{document}

\rightline{BI-TP 2012/14}

\rightline{INT-PUB-12-023}

\rightline{TUW-12-14}

\maketitle

\section{Introduction}

Experiments conducted at RHIC \cite{Tannenbaum} and at the LHC \cite{Muller} suggest that the matter produced in heavy ion collisions can be identified as strongly coupled quark-gluon plasma, behaving essentially like a strongly coupled liquid. The AdS/CFT correspondence \cite{Maldacena,Gubser,Witten}, on the other hand, is a powerful tool for studying the properties of strongly interacting, large-$N$ field theories at finite temperature and density; for recent reviews see e.g.~\cite{Wiedemann,Mateos}. In the AdS/CFT setting, thermalization on the field theory side is related to the classical gravity description of black hole formation in a 5-dimensional spacetime.

While holographic thermalization is in general a very complicated problem (discussed e.g.~in \cite{Banks,Balasubramanian,CaronHuot:2011dr}), an interesting approach was taken in \cite{Danielsson1,Danielsson2}, where the authors studied the quasistatic formation of a black hole through the gravitational collapse of a thin shell of matter in AdS space, falling in the radial direction. Amongst other things, they determined the retarded Green's function of a massive scalar field living on the boundary of AdS space, as the shell radius decreased towards the Schwarzschild radius and the field theory system approached equilibrium. In \cite{LinShuryakI,LinShuryakIII}, this picture was further extended by analyzing the correlators of metric perturbations.

In the paper at hand, we consider the $R$-current correlator in strongly coupled $\mathcal{N}=4$ supersymmetric Yang-Mills theory (SYM) in the large $N$ limit, having in mind the spectral function related to the production of dileptons at rest \cite{Huot}. On the gravity side, this amounts to introducing an effective U(1) gauge field in $\mathrm{AdS}_5$ space and determining the corresponding spectral function as well as the quasinormal spectrum. In contrast to earlier works on the subject, we want to perform the calculation in an out-of-equilibrium setting, and to this end follow references \cite{Danielsson1,Danielsson2} by introducing to the system a thin, spherical shell of matter that undergoes gravitational collapse in the fifth (radial) dimension of the AdS space. When the radius $r_s$ of the shell approaches the Schwarzschild radius $r_h$ dictated by its mass, the dual field theory system undergoes thermalization. In the present paper, we restrict ourselves to a limit where the motion of the shell can 
be considered quasistatic, and hence the production rate of dileptons is available using methods discussed e.g.~in \cite{Myers}. The (QCD version of the) quantity we compute is measurable in real world experiments and is moreover a very interesting probe of the initial stages of a heavy ion collision; for reviews, see e.g.~\cite{Srivastava,PHENIX,Blaizot} and references therein.

Our paper is organized as follows: In section \ref{matching}, we introduce the setup in $\mathrm{AdS}_5$ space by writing down the corresponding metric and discussing the matching conditions for different types of fields at the collapsing shell. In Section \ref{dilepton}, we then determine the relevant holographic Green's functions and relate them to dilepton production, while in section \ref{conclusions} we draw our conclusions. In appendix \ref{AdS3}, we finally discuss the case of a massive scalar field in 3-dimensional $\mathrm{AdS}_3$ space, and in appendix \ref{fdt} the validity of the Fluctuation Dissipation Theorem (FDT) in our setup.

After the first version of our paper was posted on the arxiv, several interesting and closely related papers on holographic thermalization appeared. In \cite{Erdmenger2}, the authors analyze the falling shell setup and manage to go beyond the quasistatic approximation in their study of unequal-time correlators. Ref.~\cite{Caceres} on the other hand studies holographic thermalization in the presence of a finite chemical potential, while ref.~\cite{Mukhopadhyay} attempts to derive a holographic FDT valid out of thermal equilibrium.

\section{Setup and matching conditions}\label{matching}

Our aim is to follow the idea of Danielson et al.~\cite{Danielsson1,Danielsson2} and describe the thermalization process in strongly coupled, large-$N$ ${\mathcal N}=4$ SYM plasma via the gravitational collapse of an infinitely thin shell of matter in a background, which is asymptotically $\mathrm{AdS}_5$. Analogously to Birkhoff's theorem, we know that outside the shell the metric is given by a black hole solution, i.e.~the Schwarzschild solution in $\mathrm{AdS}_5$; similarly, the metric inside the shell is given by the flat $\mathrm{AdS}_5$ metric. Thus, we can write the metric in our setup in the form
\begin{equation}
\label{metric} ds^2 \,=\, -f(r)dt^2 + \frac{dr^2}{f(r)}+r^2d\mathbf{x}^2 \; ,
\end{equation}
where $\mathbf{x}$ stands for the coordinates of Euclidean 3-space and we have set the curvature radius of AdS space to unity. If the shell resides at $r = r_s$, then the function $f(r)$ is given by
\be
f(r) \,=\, \left\{ \begin{array}{lr}
f_-(r) = 1+ r^2 \, ,& \mathrm{for}\;r < r_s\\
f_+(r) = 1 - \frac{m^2}{r^2}+r^2 \, ,& \mathrm{for}\; r > r_s
\end{array}\right. \; ,
\ee
where the parameter $m$ is related to the mass of the shell. To match the two (in principle unrelated) coordinate patches, we choose the obvious condition $\mathbf{x}|_+ \equiv \mathbf{x}|_-$. From this, it follows that the $r$-coordinate is continuous over the boundary, while the time coordinate $t$ has a discontinuity at the shell. Finally, in the limit where the mass of the shell is large enough so that the corresponding horizon radius $r_h\gg 1$, we can relate the parameter $m$ to $r_h$ and the Hawking temperature of the black hole $T$ through
\begin{equation}
\label{temperature} m \,=\, r_h^2 \,=\, \pi^2 T^2 \; .
\end{equation}
In the presence of a black hole, i.e.~once the thermalization process has ended, the parameter $T$ is identified with the equilibrium temperature of the field theory; in the rest of this paper, we will, however, refer to this definition of $T$ even when the shell resides at a radius $r_s>r_h$.

We are interested in finding matching conditions for various types of fields at $r=r_s$, and to this end, we need the normal vector of the shell. Let the coordinates of the shell be
\[ t \,=\, t_s(\tau)\, , \quad r \,=\, r_s(\tau) \; ,\]
where the parameter $\tau$ is the proper time of the shell. From the condition $u^\mu u_\mu = -1$, we find
\be
-f_\pm(r_s)\dot{t}_s^2+\frac{\dot{r}_s^2}{f_\pm(r_s)} \;=\; -1 \;,
\ee
where we have denoted a derivative with respect to $\tau$ by a dot; from here, it also follows that
\begin{equation}
\label{tdot} \dot{t}_{s\pm} \,=\, \frac{\sqrt{f_\pm(r_s)+\dot{r}_s^2}}{f_\pm(r_s)} \; .
\end{equation}
Next, we introduce the normal vector of the shell, $n^\mu$, and require it to be normal to the four-velocity of the shell as well as properly normalized. This produces
\ba
n_t\dot{t}_s + n_r\dot{r}_s &=& 0 \; , \\
 -\frac{n_t^2}{f_\pm(r_s)} +f_\pm(r_s)n_r^2 &=& 1 \; ,
\ea
which after some algebra leads us to the simple result
\be
[n_\mu] \,=\, (-\dot{r}_s,\dot{t}_s,0,0,0) \; .
\ee
Note that while $\dot{r}_s$ is the same on both sides of the shell, $\dot{t}_s$ is not, but has a discontinuity given by eq.~(\ref{tdot}).

To determine the trajectory of the shell, $r_s(\tau)$, one should next specify its energy content, after which one would proceed to solve its geodesic equation. While this information is in principle crucial to disentangle the time evolution of the thermalization process, we will in the present work restrict ourselves to the quasistatic limit, where the frequency of the modes we study is assumed to be much larger than the inverse time scale related to the falling of the shell. This produces a crucial technical simplification, and corresponds to a particular choice of initial conditions for the thermalization process. In section \ref{conclusions}, we will inspect the validity of this approximation, and in addition relate the value of the parameter $m$ to the characteristic time scale of the thermalization process.

\subsection{Junction condition for a scalar field}

Let us first study the behavior of a scalar field $\phi$ at the shell radius, both for simplicity and for the eventual use of the results in appendix \ref{AdS3}. Its junction condition states that the covariant derivative normal to the shell should be continuous across the shell \cite{Giddings}, i.e.
\be
n^\mu\nabla_\mu \phi \big|_- = n^\mu \nabla_\mu\phi\big|_+ \; .
\ee
Writing this out explicitly in coordinate space gives
\be
-\frac{\dot{r}_s}{f_-(r_s)}\frac{\partial \phi}{\partial t_-} \bigg|_- +\sqrt{f_-(r_s)+\dot{r}_s^2}\; \frac{\partial \phi}{\partial r} \bigg|_- = -\frac{\dot{r}_s}{f_+(r_s)}\frac{\partial \phi}{\partial t_+} \bigg|_+ +\sqrt{f_+(r_s)+\dot{r}_s^2} \;\frac{\partial \phi}{\partial r} \bigg|_+ \;,
\ee
where $t_\pm$ reminds us of the fact that the time coordinates are different on the two sides of the shell, and thus need to be evaluated separately.

The quasistatic approximation is equivalent to assuming that the timescale related to the motion of the shell is longer than any other timescale of interest. We may thus approximate $\dot{r}_s\to0$, which gives us
\be
\sqrt{f_-(r_s)}\left.\frac{\partial\phi}{\partial r}\right|_- \,=\, \sqrt{f_+(r_s)}\left.\frac{\partial\phi}{\partial r}\right|_+ \; .
\ee
Assuming further the continuity of $\phi(t)$ across the shell, $\phi_-(t_-)=\phi_+(t_+)$, and using the relation
\be\label{fm2}
\frac{dt_-}{dt_+}\,=\,\sqrt{\frac{f_+}{f_-}}\,\equiv\, \sqrt{f_m}\quad \Rightarrow \quad \int d t_+ e^{i\w_+ t_+}=\frac{1}{\sqrt{f_m}}\int dt_-e^{i\frac{\w_+ t_-}{\sqrt{f_m}}}\; ,
\ee
this leads to the identification $\w_-=\w_+/\sqrt{f_m}$ as well as to the Fourier space matching conditions
\ba\label{mscalar}
 \phi_-(\omega_-) &=& \sqrt{f_m}\phi_+(\omega_+)\, , \\
  \phi'_-(\omega_-) &=& f_m\phi'_+(\omega_+)\, ,
\ea
where the prime denotes a derivative with respect to $r$.

\subsection{Junction condition for a vector field}

To find the junction condition in the vector field case, we loosely follow the derivation given in \cite{Giddings}. We first rewrite the metric in the form
\be
ds^2 \,=\, -f\,dt^2+\frac{dr^2}{\tilde{f}_\pm} +r^2\,d\mathbf{x}^2 \; ,
\ee
where the $f$-functions are given by
\begin{eqnarray}
f &=& \left\{ \begin{array}{lr}
r^2\left( 1-\frac{r_h^4}{r_s^4} \right)&\mathrm{for}\quad r< r_s\\
r^2\left( 1-\frac{r_h^4}{r^4} \right)&\mathrm{for}\quad r > r_s
\end{array}\right.\\
\tilde{f} &=& \left\{\begin{array}{lr}
r^2&\mathrm{for}\quad r< r_s\\
r^2\left( 1-\frac{r_h^4}{r^4} \right)&\mathrm{for}\quad r > r_s
\end{array}\right. \; .
\end{eqnarray}
This is achieved by first approximating $r \gg 1$, and then rescaling time in the $r<r_s$ patch.

In the above coordinate system, the equation of motion for an electric field in the $z$ direction, $E = iF_{tz}$, is given by (see also eq.~(\ref{Maxwellequ}) of the following section)
\be\label{vectoreom}
E''+\frac{1}{2}\( \frac{\tf'}{\tf}+\frac{2}{r}+\frac{f'}{f}\)E'+\frac{\w^2}{f \tf}\, E =0 \, .
\ee
Introducing the tortoise coordinate
\be
\frac{dr_*}{dr}=\frac{1}{r\sqrt{ f\tf}}\;,
\ee
this simplifies to the form
\be
\partial^2_{r_*}E+\w^2r^2 E=0\, .
\ee
From this equation, we get
\be
\partial_{r_*}E |_-^+=-\int_{r_{*s}-\epsilon}^{r_{*s}+\epsilon}\w^2r^2E\,dr_* \, ,
\ee
where the right hand side furthermore vanishes in the limit of an infinitesimally thin shell ($\epsilon\to 0$). Going finally back to our original coordinate system and taking into account the jump in frequencies, we obtain from here a result very similar to the scalar field one,
\ba\label{matchingE}
E_-(\omega_-)&=&\sqrt{f_m}E_+(\omega_+),\\
E'_-(\omega_-)&=&f_m E'_+(\omega_+).
\ea

\section{Holographic dilepton production}\label{dilepton}

In this section, we want to determine the production rate of dileptons at rest in the holographic setup introduced in section \ref{matching}. We begin this in section \ref{greens} by discussing, how the dilepton production rate is related to the retarded Green's function of a photon field, and then proceed to write down the equations of motion for a U(1) vector field in the presence of a falling shell in section \ref{eqs}. After this, we evaluate the necessary retarded correlators in section \ref{retarded}, while section \ref{quasi} is devoted to a study of the quasinormal spectrum of the gauge field, providing another way to follow the thermalization process. In section \ref{wkb}, we finally analyze the large $|\w|$ limit of the correlator in more detail, applying to it the WKB approximation.

\subsection{Dilepton production rate from Green's functions} \label{greens}

It is a textbook exercise to show that the differential production rate of dileptons with four-momentum $Q$ is directly proportional to the photon Wightman function $\Pi^{<}_{\mu\nu}(Q)$ (see e.g.~\cite{lebellac}),
\begin{equation}
\frac{{\rm d}\Gamma}{{\rm d}^4Q}\,=\,-\frac{\alpha \eta^{\mu\nu}\Pi^{<}_{\mu\nu}(Q)}{24\pi^4Q^2},
\end{equation}
where $\alpha$ is the fine structure constant. Just as in thermal equilibrium, one can show that in the quasistatic approximation the Wightman function can be further related to the corresponding spectral density $\chi$, i.e.~the imaginary part of the retarded Green's function $\Pi_{\mu\nu}$.\footnote{This relation, a special case of the Fluctuation Dissipation Theorem, can be derived by generalizing the arguments of Ref.~\cite{Herzog} to our present case, where instead of a black hole we have a static shell sitting at a radius $r_s>r_h$. For further details, see appendix \ref{fdt}.} Setting the external three-momentum $\mathbf{q}$ to zero, which implies that the produced dileptons are at rest and that the longitudinal and transverse parts of $\Pi_{\mu\nu}$ agree,\footnote{This in particular means that it suffices to consider a single Green's function $\Pi$ from this point onwards.} we obtain 
\begin{equation}
\eta^{\mu\nu}\Pi^{<}_{\mu\nu}(\omega) \, = \, -2n(\w){\rm Im}\, \Pi^\mu_\mu(\omega) \, \equiv\, n(\w) \chi(\w) \, ,\label{fdt2}
\end{equation}
where $n(\w)$ stands for the usual Bose-Einstein distribution function. Our task thus becomes that of evaluating the retarded Green's function of a U(1) gauge field in the gravity picture.

\subsection{Equations of motion and analytic solutions} \label{eqs}

The equation of motion of a U(1) vector field in curved spacetime reads
\begin{equation}\label{Maxwellequ}
\frac{1}{\sqrt{-g}}\partial_\mu\left[\sqrt{-g} g^{\mu\rho} g^{\nu\sigma}F_{\rho\sigma}\right] = 0 \; , \quad F_{\rho\sigma}= \partial_\rho A_\sigma - \partial_\sigma A_\rho \; ,
\end{equation}
where the metric is in our case given by eq.~(\ref{metric}). Once again, we approximate $r \gg 1$ and write the mass in terms of the horizon radius $r_h$. To simplify the otherwise cumbersome expressions, we also introduce the commonly used variables
\[ u \equiv \frac{r_h^2}{r^2}\quad \mathrm{and} \quad z\equiv 1-u \; ,\]
which both vary between 0 and 1. In the following, we will repeatedly interchange between the three radial variables; it should, however, always be clear from the context, which one is being used at a given time.

The gauge invariant component of the field strength tensor we are interested in can be chosen as $F_{tz}(t,u,\mathbf{x})$. Given in terms of the Fourier components of the vector potential, $A_\mu(t,u,\mathbf{x}) = \int_Q e^{-i\omega t+i\mathbf{q}\cdot\mathbf{x}}A_\mu(\omega,u,\mathbf{q})$, this quantity reads \cite{Huot,Myers}
\begin{equation}
\label{gaugefield}
F_{tz}(\omega,u,\mathbf{q}=0) = -i\omega A_z(\omega,u) \,\equiv \, -iE(\omega,u) \; ,
\end{equation}
where $E$ stands for the electric field in the $z$ direction and we have set the three-momentum $\mathbf{q}$ to zero. A short calculation gives for the equation of motion of $E$
\begin{equation}
\label{eomEz}
\partial^2_z E + \frac{\partial_z f}{f}\partial_zE +\frac{\hat{\omega}^2}{(1-z)f^2}E = 0\;,
\end{equation}
where we have introduced the dimensionless variable
\[ \hat{\omega} \equiv \frac{\omega}{2r_h} = \frac{\omega}{2\pi T}\;.\]
This equation can be identified with the Riemann differential equation \cite{Abramowitz},
\begin{equation}
\partial_z^2E + \left[ \frac{1}{z}+\frac{1}{z-2}\right]\partial_z E + \left[ \frac{\hat{\omega}^2/2}{z} - \frac{\hat{\omega}^2/2}{z-2}\right]\frac{E}{z(z-1)(z-2)}= 0\; ,
\end{equation}
which has two linearly independent solutions,
\begin{eqnarray}\label{Esol}
&&\!\!\Ein(\omega,z) = z^{-\frac{i\hat{\omega}}{2}}(1-z)^{\frac{(1+i)\hat{\omega}}{2}}(2-z)^{-\frac{\hat{\omega}}{2}}{}_2F_1\left(1-\frac{1+i}{2}\hat{\omega},-\frac{1+i}{2}\hat{\omega};1-i\hat{\omega};\frac{z}{2(z-1)}\right) \;\;\;\;\;\; \\
&&\!\!\Eout(\omega,z) = z^{\frac{i\hat{\omega}}{2}}(1-z)^{\frac{(1-i)\hat{\omega}}{2}}(2-z)^{-\frac{\hat{\omega}}{2}}{}_2F_1\left(1-\frac{1-i}{2}\hat{\omega},-\frac{1-i}{2}\hat{\omega};1+i\hat{\omega};\frac{z}{2(z-1)}\right) \;\; \;\;\;
\end{eqnarray}
that can be expressed in terms of the hypergeometric functions ${}_2F_1$ (see also \cite{Myers}). Here, $\Ein$ is the so-called infalling solution, obeying the incoming wave boundary condition at the horizon, $z\to 0$, while $E_\textrm{out}$ satisfies the outgoing boundary condition in this limit. In the limit $z\to 0$, or $r\to r_h$, the two solutions can be approximated by the expressions
\begin{equation}
E_{\substack{\mathrm{in}\\\mathrm{out}}} (\omega,z\to0) \;\simeq\; z^{\mp i \hat{\omega}/2} \; = \; e^{\mp \frac{i\hat{\omega}}{2} \ln \left(1-\frac{r_h^2}{r^2} \right)} \; .
\end{equation}

In the presence of a black hole, the physical solution is the one obeying the infalling boundary condition, as classically nothing can escape from inside the event horizon \cite{SonStarinets}. In our case, waves can, however, be reflected from the shell and escape to the boundary, so outside the shell we must write the general solution as a linear combination of the two independent solutions (here and in the following, we use $\omega$ to denote the frequency outside the shell),
\begin{equation}
E_\mathrm{outside} (\omega,r) = c_+ \Ein(\omega,r) +c_-\Eout(\omega,r) \;.
\end{equation}
Inside the shell, $r<r_s$, the solution is on the other hand obtained from the $f\to 1$ limit of eq.~(\ref{eomEz}), giving
\begin{equation}
E_\mathrm{inside} (\omega,r) = \frac{1}{r}\left[ J_1\left(\frac{\omega_\mathrm{inside}}{r}\right) + i Y_1 \left(\frac{\omega_\mathrm{inside}}{r}\right) \right] \stackrel{r\to0}{=} \frac{1}{r}\sqrt{\frac{r}{\omega_\mathrm{inside}}}e^{i \omega_\mathrm{inside}/r} \; ,
\end{equation}
where again $\omega_\mathrm{inside} = \omega/\sqrt{f_m}$, with $f_m$ defined in eq.~(\ref{fm2}). The matching conditions of eq.~(\ref{matchingE}) then lead to the result
\begin{eqnarray}
\label{Cmp}
\frac{c_-(r_s)}{c_+(r_s)} = - \frac{\Ein(\omega,r)\partial_r E_\mathrm{inside}(\w,r)-\sqrt{f_m} E_\mathrm{inside}(\w,r) \partial_r \Ein(\omega,r)}{\Eout(\omega,r)\partial_r E_\mathrm{inside}(\w,r)-{\sqrt{f_m}} E_\mathrm{inside}(\w,r)\partial_r \Eout(\omega,r)}\Bigg|_{r=r_s} \;,
\end{eqnarray}
which we will use frequently below. For real values of $\omega$, this ratio satisfies the relation
\ba
\frac{c_-(r_s)}{c_+(r_s)}\Bigg|_{\omega\to -\omega} =\;\, \frac{c^*_-(r_s)}{c^*_+(r_s)} \;  \label{comprel}
\ea
that will prove rather useful in the following.

\subsection{The retarded Green's function and spectral density} \label{retarded}

The retarded correlator (Green's function) of an operator living on the boundary of AdS space can be determined using the standard AdS/CFT prescription, developed and thoroughly explained in \cite{SonStarinets, Kovtun}. The wave function of the corresponding bulk field is first written in terms of a power series expansion around the singular point(s) of its equation of motion (in our case $E_\mathrm{outside}$ at $u=0$),
\be
\label{boundexp}
E_\mathrm{outside}(\omega, u) \stackrel{u \to 0}{=} \mathcal{A}(\omega)u^{\Delta_-}\left[ 1 + \ldots\right] + \mathcal{B}(\omega)u^{\Delta_+}\left[1 + \ldots \right],
\ee
where $\Delta_\pm$ are the related characteristic exponents. For $E_\mathrm{outside}$, we have $\Delta_+=1$ and $\Delta_-=0$, and hence the solution takes the form
\be\label{boundexp1}
E_\mathrm{outside}(\omega, u) \stackrel{u \to 0}{=} \mathcal{A}(\omega)\left[ 1+h(\w) u\ln(u) + \ldots\right] + \mathcal{B}(\omega)\left[u + \ldots \right]\;.
\ee
The retarded Green's function of the boundary field is then given by the expression
\be\label{Gret}
\Pi (\omega) = -\frac{N_c^2T^2}{8}\frac{\mathcal{B}(\omega)}{\mathcal{A}(\omega)} \; ,
\end{equation}
which can equivalently be expressed in terms of its thermal limit times a correction function,
\begin{equation}
\Pi(\omega) = \Pi^\mathrm{thermal}(\omega)\times H(\omega,r_s) \; ,
\end{equation}
where we have explicitly indicated the fact that the function $H$ depends on the radius of the shell, and where obviously $H\to 1$ as $r_s\to r_h$.

For our electric field, the thermal limit of the retarded Green's function can be read off from the literature (see e.g.~\cite{Myers}),
\begin{eqnarray}
\Pi^\mathrm{thermal}(\omega) &=& -\frac{N_c^2T^2}{8} \frac{\mathcal{B}_\mathrm{in}}{\mathcal{A}_\mathrm{in}}\nonumber\\
&=& \frac{N_c^2T^2}{8}\left\{ i\hat{\omega} + \hat{\omega}^2\left[ \psi\left( \frac{1-i}{2}\hat{\omega}\right)+\psi\left(-\frac{1+i}{2}\hat{\omega}\right)+\ln 2+ 2\gamma -1\right]\right\} \, , \label{Gtherm}
\end{eqnarray}
while the function $H$ obtains the form
\be
 H(\omega,r_s)=
\frac{1+\frac{c_-}{c_+}\Big|_{r_s}
\frac{\mathcal{B}_\mathrm{out}}{\mathcal{B}_\mathrm{in}}}{1+\frac{c_-}{c_+}\Big|_{r_s}\frac{\mathcal{A}_\mathrm{out}}{\mathcal{A}_\mathrm{in}}} \; \label{H}.
\ee
Here, $\psi$ is the logarithmic derivative of the gamma function, while $\gamma\approx 0.5772$ is the Euler-Mascheroni constant. The imaginary part of the thermal Green's function is identified with the spectral function, which represents the thermal production rate of dileptons at rest. It is a smooth function of $\hat{\omega}>0$ and reads
\begin{equation}
\chi^\mathrm{thermal} (\w)=-2\,\mathrm{Im}\,\Pi^\mathrm{thermal} (\omega) = \frac{N_c^2T^2}{4}\frac{\pi \hat{\omega}^2 \sinh (\pi \hat{\omega})}{\cosh( \pi\hw) - \cos( \pi\hat{\omega})} \; .
\end{equation}

\begin{figure}
\centering
\includegraphics[width=10cm]{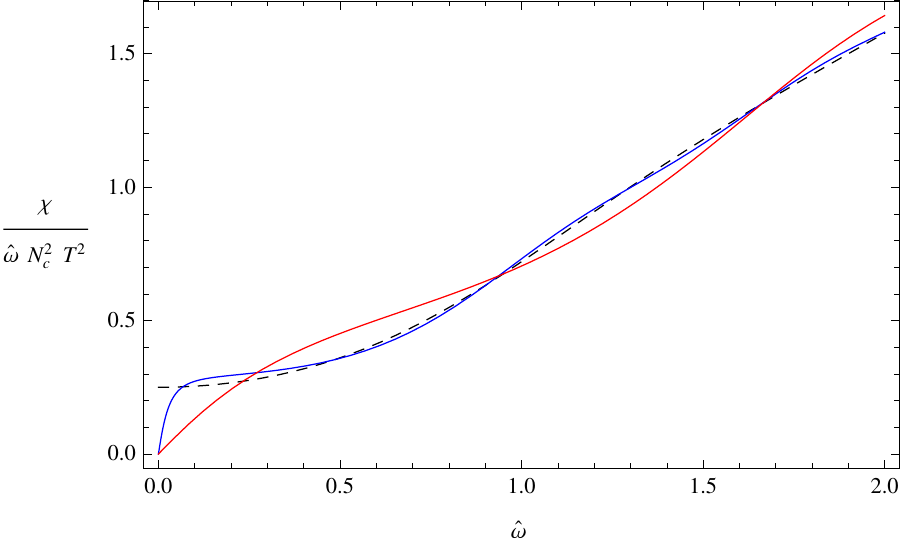}
\caption{The spectral density $\chi(\w,r_s)$ as a function of $\hat{\w}$ for $r_s=r_h=5$ (thermal case; dashed, black curve), $r_s=5.01$ (blue) and $r_s=5.5$ (red). Note that we use units, in which the radius of AdS space has been set to unity.}
\label{rhosmall}
\end{figure}

Next, let us study the Green's function outside of thermal equilibrium. From an expansion of the infalling solution (\ref{Esol}), one can read off the coefficients in eq.~(\ref{boundexp1}) as
\begin{eqnarray}
\mathcal{A}_\mathrm{in} &=& \frac{\Gamma(1-i\hat{\omega})\exp \left[ -\frac{1+i}{2}\hat{\omega} \ln 2\right]}{\Gamma\left( 1-\frac{1+i}{2}\hat{\omega}\right)\Gamma\left(1+\frac{1-i}{2}\hat{\omega}\right)}\;,\\
\mathcal{B}_\mathrm{in} &=& \mathcal{A}_\mathrm{in}(\omega) \left\{ -i\hat{\omega}-\hat{\omega}^2\left[ \psi\(\frac{1-i}{2}\hat{\omega}\)+\psi\(-\frac{1+i}{2}\hat{\omega}\)+\ln 2 + 2\gamma -1\right]\right\} \; ,
\end{eqnarray}
while their outgoing counterparts are obtained through the replacements
\be
\mathcal{A}_\mathrm{out}(\omega)\, =\, \mathcal{A}_\mathrm{in}(i\hat{\omega}\to-i\hat{\omega})\, ,\quad \mathcal{B}_\mathrm{out}(\omega)\, =\, \mathcal{B}_\mathrm{in}(i\hat{\omega}\to-i\hat{\omega})\, .
\ee
From eq.~(\ref{Cmp}), we further see that our results fulfill an important consistency condition: As the shell approaches the horizon, i.e.~$r_s\to r_h$ and $f_m\rightarrow 0$,
\be
\lim_{r\rightarrow r_s}\frac{c_+(r)}{c_-(r)}=-i\,  2^{2+i\hat{\w}}\,\hat{\omega} \(\frac{r_s}{r_h}-1 \)^{-\frac{1}{2}+i\hat{\omega}}
\to \infty\, \label{cpmapp}
\ee
and we recover the thermal correlator. In fig.~\ref{rhosmall}, we plot the spectral density $\chi(\w,r_s)=-2\,\mathrm{Im}\,\Pi (\omega)$ as a function of frequency both for the thermal and non-thermal cases.

\begin{figure}
\centering
\includegraphics[width=10cm]{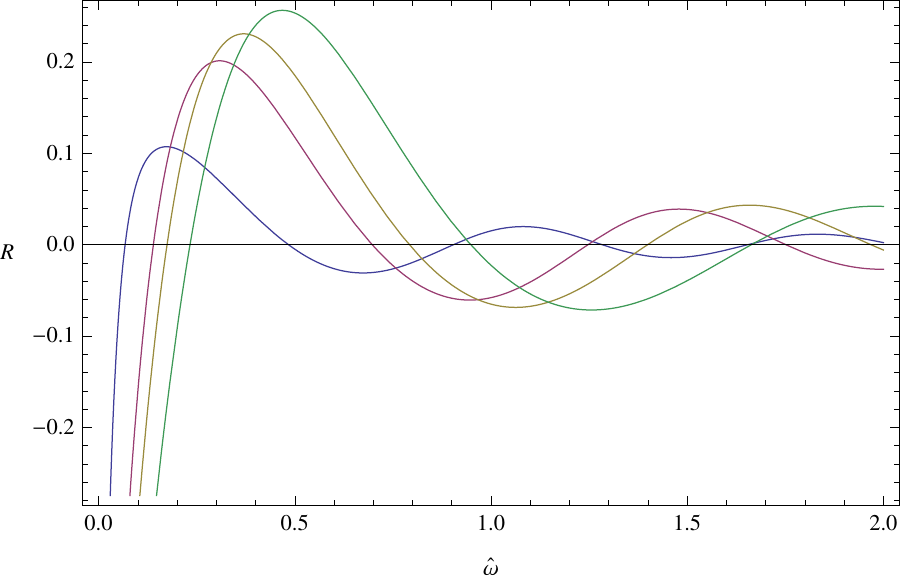}
\caption{The function $R(\w,r_s)$ shown for various values of $r_s$ ($r_h=5$). In order of increasing amplitude, the curves correspond to $r_s=5.01,\,5.1,\;5.2,\;5.5$.}
\label{Rsmall2}
\end{figure}

In order to further study the thermalization process, we next follow Lin and Shuryak \cite{LinShuryakIII} and plot in fig.~\ref{Rsmall2} the relative deviation of the spectral density from its thermal limit,
\be
\label{R}
R(\w,r_s)=\frac{\chi(\w,r_s)-\chi^\mathrm{thermal}(\w)}{\chi^\mathrm{thermal}(\w)},
\ee
for different values of $r_s$. As the shell approaches the horizon, the oscillations visible in the function $R$ increase in frequency and decrease in amplitude until they finally vanish, as the medium thermalizes. For purposes of clarity, we have chosen not to display here the entire range of $R$, which approaches -1 as $\w\to 0$ for all values of $r_s$.\footnote{Note that this descrease in $R$ occurs for values of $\omega$ that are outside of the region of validity of the quasistatic approximation, cf.~section \ref{conclusions}. See also \cite{Rebhan} for discussion of a similar behavior of the spectral density, observed in a somewhat different setting.} From figs.~\ref{rhosmall} and \ref{Rsmall2}, we further see that as expected, the approach of the system towards thermal equilibrium is of the ``top-down'' type: The energetic modes, i.e.~high mass dileptons, equilibrate first. This feature of strong coupling thermalization was already observed e.g.~in \cite{Balasubramanian,LinShuryakIII} (for a discussion 
of thermalization in a different context, see also \cite{CheslerYaffe} and references therein). It is to be contrasted to the behavior of weakly coupled ``bottom  up'' scenarios such as \cite{Baier,Kurkela}, in which the soft modes thermalize before the hard ones and one would naively expect the amplitude of $R$ to increase as a function of $\hat{\omega}$.

\subsection{Quasinormal modes} \label{quasi}

Analyzing the quasinormal spectrum of a field living in the bulk --- available from the poles of its retarded Green's function --- offers a convenient way to study the effects of the plasma on the corresponding field theory excitation, as well as the thermalization process itself. On the field theory side, the spectrum gives the dispersion relation of the mode in question, which in the limit $r_s\to r_h$ should approach the thermal result. A departure from equilibrium is expected to lead to deviations from this spectrum, as well as to the appearance of altogether new poles in the correlator. At zero three-momentum, these poles have the generic form
\be
\w_n=M_n-\frac{1}{2}\Gamma_n,
\ee
where the real part $M_n$ is identified with the mass of the resonance and the imaginary part $\Gamma_n$ with the corresponding decay rate.

Recall that we can write the full retarded correlator in a factorized form
\be
\Pi(\w)=\Pi^\textrm{thermal}(\w)\times H(\w,r_s),
\ee
where $\Pi^\textrm{thermal}$ and $H$ are given by eqs.~(\ref{Gtherm}) and (\ref{H}), respectively. The thermal correlator is known to possess a quasinormal spectrum with poles located at \cite{Myers} (see also \cite{Berti,Landsteiner})
\begin{equation}
\label{QNMtherm}
\omega \,=\, 2\pi Tn(\pm 1-i)\,=\, 2r_h n(\pm 1-i) \; , \quad n = 0,1,\ldots
\end{equation}
which we expect to be shifted when $r_s\neq r_h$. The approach of the out-of-equilibrium spectrum towards the thermal one can be analyzed using eq.~(\ref{cpmapp}): We see that for all modes satisfying the condition Im$(\w)>-r_h$, the ratio $c_-/c_+$ approaches 0 as $r_s\to r_h$, and thus the poles approach the thermal form. For Im$(\w)<-r_h$, the quasistatic approximation on the other hand appears to break down, and hence the thermal modes with $n>0$ will not be seen in the $r_s\to r_h$ limit.

\begin{figure}
\centering
\includegraphics[width=10cm]{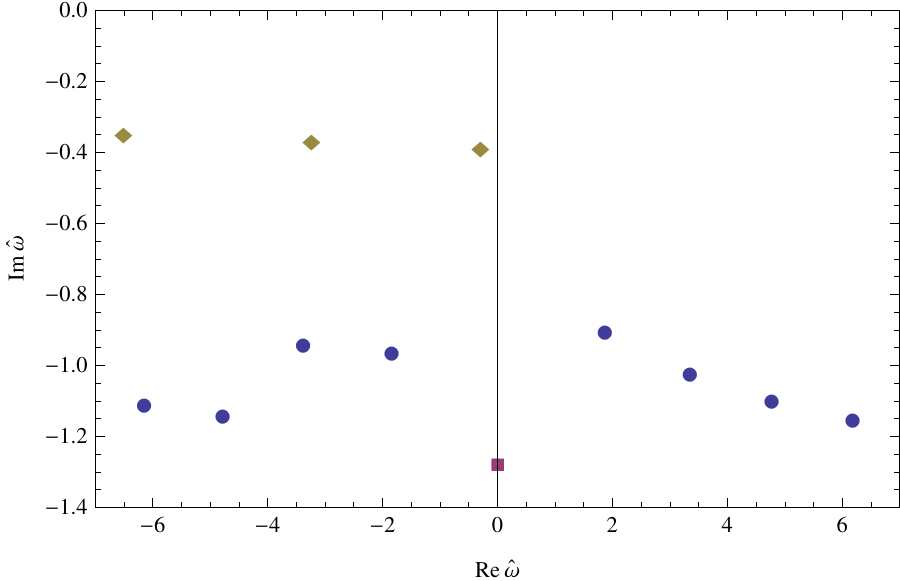}
\caption{The poles of the full retarded correlator for $r_s=5.5$ (with $r_h=5$).}
\label{QNMstatic}
\end{figure}

\begin{figure}
\centering
\includegraphics[width=10cm]{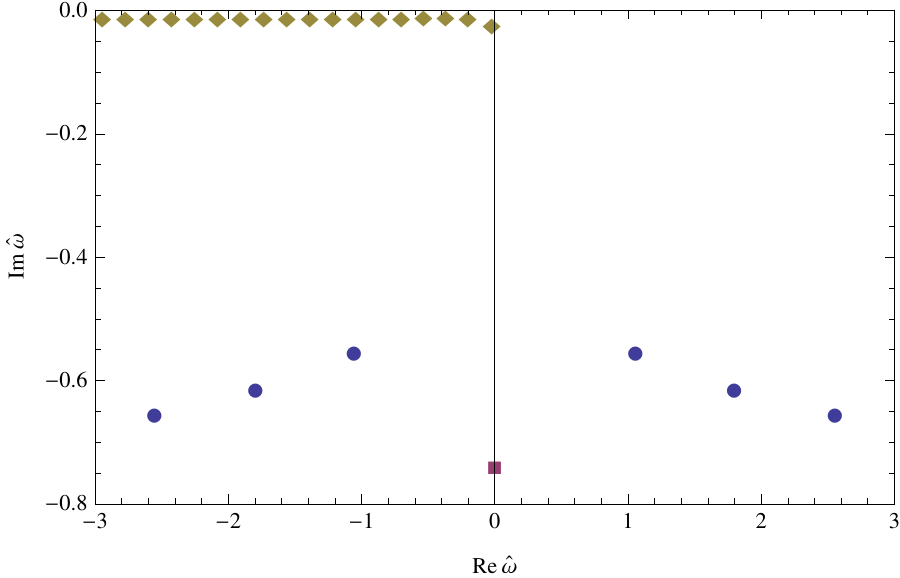}
\caption{The poles of the full retarded correlator for $r_s=5.01$ (with $r_h=5$).}
\label{QNMstatic2}
\end{figure}

In figs.~\ref{QNMstatic} and \ref{QNMstatic2}, we show the poles of the full retarded Green's function $\Pi$ for $r_s=5.5$ and $r_s=5.01$ (again with $r_h=5$), displaying only the lower complex halfplane, as the retarded correlator is analytic on the upper half. We observe that the poles of $\Pi^\textrm{thermal}$ have disappeared, as they get canceled by the zeros of $H$. At the same time, three new types of poles have appeared --- a structure that is intimately connected with the matching conditions, as they enter in the ratio of eq.~(\ref{Cmp}).

\begin{figure}
\centering
\includegraphics[width=10cm]{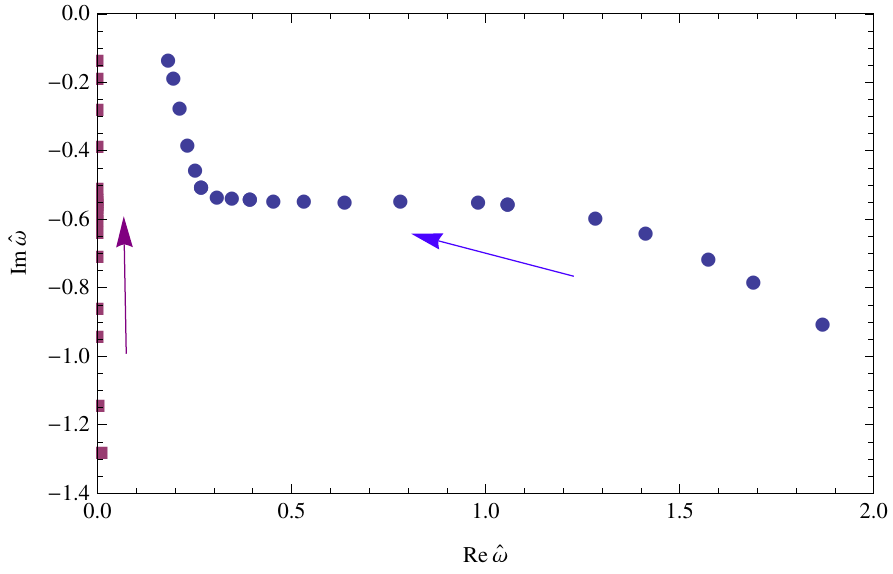}
\caption {The flow of the red and lowest blue pole of the retarded correlator on the complex $\hat{\w}$ plane, as the shell approaches the horizon at $r_h=5$. We start with the shell being located at $r_s=5.5$ and follow the flow of the poles until $r_s-r_h=10^{-15}$.}
\label{QNM}
\end{figure}

The tower of poles denoted by the blue circles are similar to the ones found in \cite{Danielsson2}, and in the limit of $r_s\rightarrow r_h$ one can understand their flow analytically. Using the approximation of eq.~(\ref{cpmapp}) together with
\ba
\frac{\mathcal{A}_\mathrm{out}}{\mathcal{A}_\mathrm{in}}\approx -ie^{2i\hat{\omega}(\ln\,2+\pi/4)},
\ea
we obtain the simple equation
\ba
\frac{r_h}{2}\sqrt{\frac{r_s}{r_h}-1}\,e^{i\omega t_\textrm{echo}}&=&-\omega,
\ea
where we have defined a parameter (the physical meaning of which will become clear later)
\be
t_\textrm{echo} = \frac{1}{2r_h}\bigg[\frac{\pi}{2}+\ln\,2-\ln\(\frac{r_s}{r_h}-1\)\bigg].
\ee
The solutions of this equation clearly coincide with the blue circular poles shown in figs.~\ref{QNMstatic} and \ref{QNMstatic2}. As $r_s \rightarrow r_h$ (see fig.~\ref{QNM}), the flow of the lowest pole approaches $\w=0$, indicating a branch point at this value \cite{Danielsson1}.

Next, let us look at the series of poles on the lower left quadrant of the complex plane, denoted by the brown diamonds in figs.~\ref{QNMstatic} and \ref{QNMstatic2}. As is evident from the figures, these poles have the property that they rapidly move towards the origin as $r_s\rightarrow r_h$. Indeed, expanding $c_-/c_+$ in powers of $r_s-r_h$ while keeping $\tilde{\omega}\equiv \omega/\sqrt{r_s/r_h-1}$ fixed and using the fact that as $\omega\rightarrow 0$, $\mathcal{A}_\mathrm{out}/\mathcal{A}_\mathrm{in}\rightarrow 1$, we obtain the simple result that the poles of the Green's function are given by the zeros of the Hankel function of the first kind,
\ba
{\rm H}_1^{(1)}\(\tilde{\omega}/2\)&=&0.
\ea
A simple numerical exercise indeed verifies this analytic result in the $r_s\rightarrow r_h$ limit.

Finally, we have the single pole denoted by the red square, whose flow exhibits a qualitative change when $\mathrm{Im}\, \omega$ reaches the value $-r_h$; cf.~fig.~\ref{QNM}. For $\mathrm{Im}\, \omega < - r_h$, its location can be described by the equations $\mathrm{Re}\, \omega = 0$ and
\be
\mathrm{Im}\, \omega = - \frac{r_h}{2^{1 - \mathrm{Im} \,\omega/(2 r_h)} }
\, \(\frac{r_s}{r_h} -1\)^{(1 - \mathrm{Im} \,\omega/r_h)/2} \, \frac{\mathcal{A}_\mathrm{out}}{\mathcal{A}_\mathrm{in}}~,
\ee
where $\frac{\mathcal{A}_\mathrm{out}}{\mathcal{A}_\mathrm{in}}$ as a function of $\mathrm{Im}\, \omega$ is real and $O(1)$ in the region of interest. However, when $r_s$ gets exponentially close to the horizon, the pole starts to approach the origin as
\be
\mathrm{Im}\, \omega = - \frac{r_h}{2} \Big(\frac{r_s}{r_h} -1\Big)^{1/2} > -r_h\, .
\ee
It would be tempting to relate this behavior to that of a diffusive pole, appearing at finite external three-momentum and reflecting the diffusive relaxation of large scale charge density fluctuations around thermal equilibrium \cite{Kovtun}. So far we have, however, not implemented $\mathbf{q}\neq 0$ in our calculation, and hence postpone this study for the future.

\subsection{The WKB approximation} \label{wkb}

So far, we have mainly studied the spectral function and quasinormal spectrum of our gauge field for small or moderate frequencies. In the opposite limit of very large $|\w|$, it is possible to gain analytic understanding of the solutions, as one may solve the retarded correlator in the WKB approximation, denoting $E(u) = 1/\sqrt{1-u^2} Z(u)$ and transforming eq.~(\ref{eomEz}) into a Schr\"{o}dinger type form
\be\label{eomWKB}
\partial^2_u Z(u) +\lk\frac{\hat{\w}^2+u}{u(1-u^2)^2} \rk Z(u)=0 \; .
\ee
The WKB solution of this equation is given by \cite{LinShuryakIII, Srivastava}
\be
\label{WKBsolution}
Z^\mathrm{WKB}_\pm(u) = S(u)^{-1/2}\exp\lk\pm i \hat{\w} \int_0^u S(u) du \rk,
\ee
where $S(u) = \sqrt{\frac{1}{u(1-u^2)^2}}$.

The full solution for $Z$ can again be written as a linear combination of infalling and outgoing modes,
\[ Z = c_+Z_\mathrm{in} + c_-Z_\mathrm{out} \; .\]
Close to the horizon, i.e.~at $u \to 1$, eq.~(\ref{eomWKB}) reduces to
\begin{equation}
Z''(u) +\frac{\hat{\w}^2 +1}{4 (1-u)^2}Z(u) \,=\, 0\, ,
\end{equation}
the solutions of which can be written as
\begin{equation}\label{Zpm}
Z_{\substack{\mathrm{in}\\\mathrm{out}}}(u) = (1 -u)^{(1 \mp i \hat{\w})/2}\;.
\end{equation}
Taking the $u\to1$ in eq.~(\ref{WKBsolution}), we on the other hand see that
\begin{equation}
Z^\mathrm{WKB}_\pm(u) \stackrel{u\to1}{\to} \sqrt{2} (1-u)^\frac{1\mp i\hat{\omega}}{2} \exp \lk \pm i\hat{\omega}\left( \frac{\pi}{4}+\ln 2\right)\rk \; ,
\end{equation}
implying that the two WKB solutions $Z^\mathrm{WKB}_\pm$ can directly be identified as infalling and outgoing modes. This leads us to the compact result
\begin{equation}
Z^\mathrm{WKB}(u) = \frac{c_+}{\sqrt{2}}\exp\left[-i\hat{\omega}a_0\right]Z^\mathrm{WKB}_+(u) + \frac{c_-}{\sqrt{2}}\exp\left[i\hat{\omega}a_0\right]Z^\mathrm{WKB}_-(u) \; ,
\end{equation}
where $a_0 = \pi/4 + \ln 2$.

Next, we look at the opposite limit of $u\to0$, in which we wish to again relate the WKB solutions to the exact ones. In this limit, eq.~(\ref{eomWKB}) reduces to
\begin{equation}
Z''(u)+\frac{\hat{\omega}^2}{u}Z(u)=0 \; ,
\end{equation}
which has the standard solutions \cite{Abramowitz}
\begin{equation} \label{besselsolutions}
Z_J(u) = \sqrt{u}J_1(2\hat{\omega}\sqrt{u}) \quad \mathrm{and} \quad Z_Y(u) = \sqrt{u} Y_1(2\hat{\omega}\sqrt{u}) \; .
\end{equation}
The Bessel functions appearing here have the asymptotic limits
\[ J_1(z) \to \sqrt{\frac{2}{\pi z}} \cos \left( z-\frac{3}{4}\pi\right), \quad Y_1(z) \to \sqrt{\frac{2}{\pi z}} \sin \left( z-\frac{3}{4}\pi\right) \; ,\]
which can be used to derive the result
\begin{eqnarray}
Z^\mathrm{WKB}(u) &=& c_+e^{-i\hat{\omega}a_0}\sqrt{\hat{\omega}u}\left[ J_1(2\hat{\omega}\sqrt{u}) + i Y_1(2\hat{\omega}\sqrt{u}) \right]e^{\frac{ 3 i \pi}{4}} \nonumber\\
&&+ c_-e^{i\hat{\omega}a_0}\sqrt{\hat{\omega}u}\left[ J_1(2\hat{\omega}\sqrt{u}) - i Y_1(2\hat{\omega}\sqrt{u}) \right]e^{-\frac{ 3 i \pi}{4}} \; .
\end{eqnarray}
Recalling finally the relation between $Z$ and $E$, and using the definition of the Hankel functions, $H_n^{(1)} \equiv J_n+i Y_n$,  $H_n^{(2)} \equiv J_n-i Y_n$, we arrive at the WKB approximation of $E$
\begin{equation}
E^\mathrm{WKB}(u) = \sqrt{\frac{u}{1-u^2}}\left[H^{(1)}_1(2\hat{\omega}\sqrt{u}) + i\frac{c_-}{c_+}e^{2i\hat{\omega}a_0} H^{(2)}_1(2\hat{\omega}\sqrt{u}) \right]\; .
\end{equation}

\begin{figure}
\centering
\includegraphics[width=10cm]{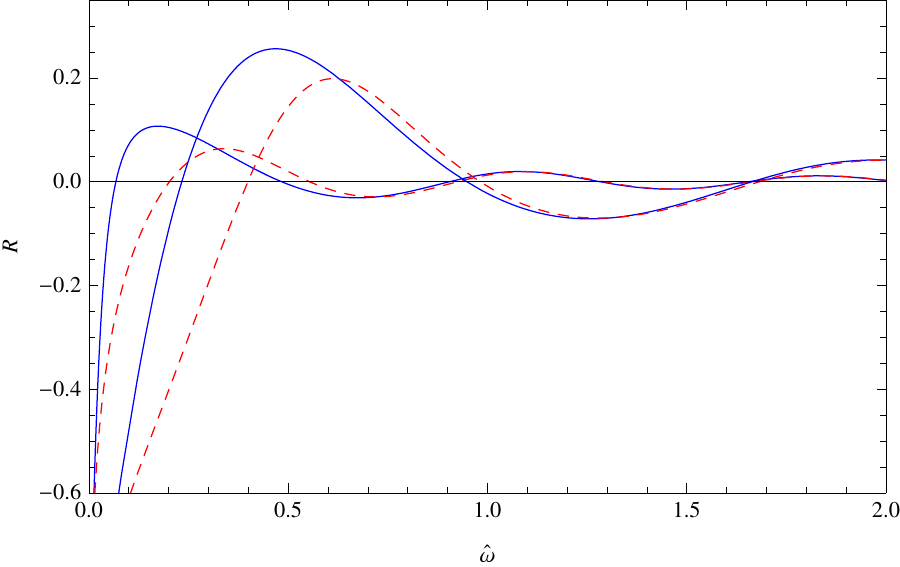}
\caption {The behavior of the functions $R(\omega)$ (blue curve) and $R^\mathrm{WKB}(\w)$ (dashed red) of eq.~(\ref{R}) for $r_s=5.01$ (smaller amplitude) and $5.5$, with $r_h=5$. }
\label{Rsmall}
\end{figure}

To arrive at an expression for the desired retarded Green's function on the field theory side, we still need to expand the WKB solution near $u = 0$ to obtain an expression similar to eq.~(\ref{boundexp1}),
\begin{eqnarray}
E^\mathrm{WKB}(u) &=& \frac{-i}{\pi \hat{\omega}\sqrt{u}}\(1-i \frac{c_-}{c_+}e^{2i\hat{\omega}a_0}\) + u \ln u \,(\ldots) \nn\\
&& + u \left[ \frac{i\hat{\omega}}{\pi}\left( -1 + 2\gamma -i \pi + 2 \ln \hat{\omega}\right)
-i \frac{c_-}{c_+}\frac{i\hat{\omega}e^{2i\hat{\omega}a_0}}{\pi}\left( -1 + 2\gamma + i\pi + 2 \ln \hat{\omega}\right)\right] \nn\\
&& + \mathcal{O}(u^2)\;.
\end{eqnarray}
From here, we can finally identify $\mathcal{A}$ and $\mathcal{B}$ and use eq.~(\ref{Gret}) to evaluate
\begin{equation}\label{GWKB}
\Pi^\mathrm{WKB}(\omega) = -\frac{N_c^2T^2}{8}\frac{\mathcal{B}}{\mathcal{A}}
= \frac{N_c^2\hat{\omega}^2}{32\pi^2}\left[2 \ln \hat{\omega} -1 + 2\gamma -i\pi\frac{1+i\frac{c_-}{c_+}e^{2i\hat{\omega}a_0}}{1-i\frac{c_-}{c_+}e^{2i\hat{\omega}a_0}} \right]\;,
\end{equation}
where the ratio $c_-/c_+$ is given in eq.~(\ref{Cmp}) and the contribution proportional to $\hat{\omega}^2 \ln \hat{\omega}$ can be identified with the $T=0$ Green's function \cite{SonStarinets}.  It is remarkable that this result has an identical structure with the retarded correlator for a scalar field, cf.~eq.~(48) of \cite{LinShuryakIII}.

Finally, we note that the large $\w$ limit of eq.~(\ref{H}) agrees with the WKB approximation, as
\be
\frac{\mathcal{A}_{out}}{\mathcal{A}_{in}}\, \stackrel{\hw\gg  1}{\longrightarrow}\, -i e^{2 i \hw a_0} ,
\ee
an effect demonstrated in fig.~\ref{Rsmall}. From the phase difference of infalling and outgoing waves near the horizon one furthermore finds the result
\be
\frac{c_-}{c_+} e^{2i\hat{\omega}a_0}
\approx \exp\lk -i\hat{\omega}\ln\(1-\frac{r_h^2}{r_s^2}\) +2i \hat{\w}a_0\rk \, ,
\ee
which allows one to derive an ``echo time'' in the limit $r_s \to r_h$ \cite{LinShuryakIII} ,
\be
t_\textrm{echo} = \frac{1}{2r_h}\lk-\ln\(1- \frac{r_h^2}{r_s^2}\) + 2 a_0\rk \; .
\ee
This quantity has a physical interpretation as the time it takes for a wave to travel back and forth the WKB potential, and can be identified as the frequency of the oscillations witnessed in the function $R(\w,r_s)$.

\section{Conclusion}\label{conclusions}

In the paper at hand, we have used the AdS/CFT correspondence to compute the retarded Green's function of a U(1) gauge field immersed in strongly coupled, thermalizing ${\mathcal N}=4$ Super Yang-Mills plasma. On the gravity side, thermalization was modeled via the gravitational collapse of an infinitesimally thin spherical shell, which was treated in a quasistationary approximation, valid in the last stages of the collapse. The goal of our work was to study the production of dileptons in an out-of-equilibrium setting --- a quantity directly proportional to the imaginary part of the retarded correlator as long as the Fluctuation Dissipation Theorem is valid. In addition, we studied the flow of the poles of the corresponding Green's function, offering an alternative way to follow the thermalization process.

The main results of our work are visible from the figures of section \ref{retarded}. The relative deviation of the spectral density from its thermal limit, defined in eq.~(\ref{R}) and displayed in fig.~\ref{Rsmall2}, is seen to exhibit fluctuations that increase in frequency and decrease in amplitude as the shell approaches the horizon and thermal equilibrium is reached. We also observe that the thermalization process is of the top-down type, as the modes with higher energy thermalize first. At the same time, figures \ref{QNMstatic}, \ref{QNMstatic2} and \ref{QNM} display the flow of the poles of the retarded Green's function as thermalization proceeds, giving the masses and decay widths of the excitations in the system. Three different types of modes were identified from these figures, and their physical interpretation discussed.

All calculations presented in this paper were performed in the quasistatic approximation, where the motion of the falling shell was assumed to be slow in comparison with the relevant time scales associated with the physical processes studied. To quantify this statement, let us note that a characteristic time scale of equilibration can be obtained from the functional dependence of the shell radius $r_s$ on $t$. Solving $r_s$ from the $ds^2 \approx 0$ limit of eq.~(\ref{metric}), we obtain an exponentially slow approach towards the horizon \cite{Danielsson2},
\be
\frac{r_s(t) - r_h}{r_h}  = \text{e}^{-t/\tau}  , \label{rstau}
\ee
where $\tau = 1/(4 \pi T)$. Denoting by $\w$ again the frequency variable in the Green's functions, we obtain from here the consistency condition
\be
\frac{|\dot{r}_s(t)|}{r_s(t)}\; \ll \w \quad \Rightarrow \quad \frac{\text{e}^{-t/\tau}}{\tau} \; \ll \; \w\, ,
\ee
which is observed to be satisfied for sufficiently high frequencies \textit{or} late times. Furthermore, plugging in eq.~(\ref{rstau}) $T \gtrsim 200$ MeV, one obtains the thermalization time scale $\tau \lesssim 0.1$ fm, which one might compare with the typical production time of dileptons with mass/energy $Q=\w$ larger than 5 GeV, $\tau_p \lesssim 0.04$ fm. It appears that dilepton pairs produced early on have a reasonable chance to escape the system while it is still out of thermal equilibrium.

While already our present results have provided interesting signatures of dilepton production, we would like to emphasize that this work should be considered only a first step towards a more complete holographic treatment of the phenomenon in an out-of-equilibrium setting. The next goal would clearly be to try to relax the quasistatic approximation in the present setup, taking a step towards fully dynamic thermalization (for closely related work, see e.g.~\cite{Erdmenger,CheslerTeaney}). Another assumption made in the current work that would be nice to relax is that of working in a spatially homogenous and isotropic background: The holographic system closest to the initial conditions of a heavy ion collision is clearly that of two colliding shock waves \cite{CheslerYaffe2,Wu}. In addition to this, it would of course be highly interesting to determine other, related quantities, such as prompt photon production, in a thermalizing plasma \cite{future}.

\section*{Acknowledgments}
We thank Ville Ker\"anen for invaluable advice concerning the evaluation of different holograhpic Green's functions, as well as Janne Alanen, Paul Chesler, Johanna Erdmenger, Esko Keski-Vakkuri, Shu Lin and Dominik Steineder for useful discussions. S.S.,~O.T.~and A.V.~were supported by the Sofja Kovalevskaja program of the Alexander von Humboldt Foundation, and S.S.~additionally by the START project Y435-N16 of the Austrian Science Fund (FWF). A.V.~would in addition like to acknowledge the Institute for Nuclear Theory (Seattle) and its program \textit{Gauge Field Dynamics In and Out of Equilibrium} for hospitality during a time when important progress was made in the project.



\begin{appendix}
\section{Massive scalar field}\label{AdS3}

In this appendix, we consider the case of a scalar field with mass $m$, frequency $\w$ and vanishing three-momentum in AdS$_3$ space, in order to make sure that our results for this somewhat simpler system agree with those of \cite{Danielsson1,Danielsson2}. Introducing the variable $u=r_h^2/r^2$, the equation of motion for the field $\phi$ reads \cite{KeskiVakkuri,Birmingham}
\be
u(1-u)\phi''(\w,u)-u\phi'(\w,u)+\lk\frac{\hat{\w}^2}{1-u}-\frac{\hat{m}^2}{u}\rk\phi(\w,u)=0\;,
\ee
where $\hat{w}\equiv\w/(2r_h)$ and $\hat{m}^2\equiv m^2/4$. Inside the shell, $r \leq r_s$, the frequency reads again $\w_{in}=\w/\sqrt{1-r_h^2/r_s^2}$, and the solution is given by a linear combination of Bessel functions $J_{\pm\nu}$ \cite{Danielsson1},
\be
\phi_\mathrm{inside}(\w,r)=\frac{1}{r}\lk J_\nu\(\frac{\w_{in}}{r}\)-e^{i\pi\nu}J_{-\nu}\(\frac{\w_{in}}{r}\)\rk\;,
\ee
with $\nu\equiv\sqrt{1+m^2}$.

\begin{figure}
\centering
\includegraphics[width=10cm]{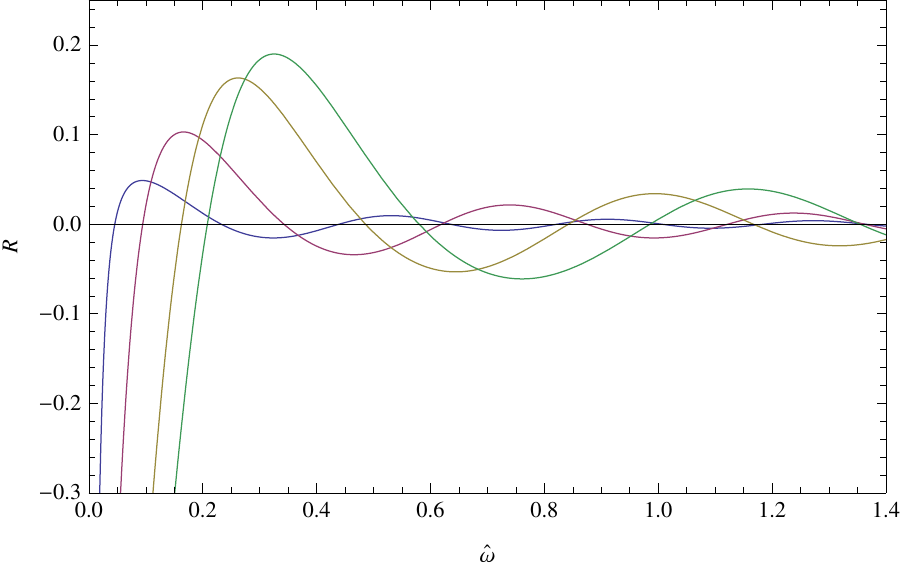}
\caption {The behavior of $R(\w,r_s)$ for $r_h=5$. In order of increasing amplitude, the curves correspond to $r_s=5.001,\,5.01,\;5.05,\;5.1$.}
\label{Einstein}
\end{figure}

For the exterior solution at $r>r_s$, we again take a linear combination of infalling and outgoing waves,
\be\label{phicomb}
\phi_\mathrm{outside}(\w,r)=c_+\phi_\mathrm{in}(\hw,r)+c_-\phi_\mathrm{out}(\hw,r)\;,
\ee
where we have defined
\be\label{phioutside}
\phi_{\substack{\mathrm{in}\\\mathrm{out}}}(\hw,r)=z^{\pm i\hw}(1-z)^{\beta_-}{}_2 F_1\(\pm i\hw+\beta_-;\pm i\hw+\beta_-;1\pm 2i \hw;z\)\, ,
\ee
and $\beta_\pm\equiv\Delta_\pm/2$ with $\Delta_\pm=1\pm\sqrt{1+m^2}=1\pm\nu$. The hypergeometric function ${}_2  F_1$ appearing in this expression is regular in the limit $r\to r_h$, in which we obtain the asymptotic behavior
\be
\phi_{\substack{\mathrm{in}\\\mathrm{out}}}(\w,{r\to r_h})\;\to\; e^{-i\w t}e^{\mp\frac{i\w}{4\pi T}\ln(\frac{r^2-r_h^2}{r_h^2})}\; . \label{phihor}
\ee
The coefficients in eq.~(\ref{phicomb}) are finally fixed by the matching conditions of eq.~(\ref{mscalar}), and hence in the full retarded Green's function on the boundary ($r\to \infty$)
\be
G_\textrm{R}(\w,r_s)=G^\mathrm{thermal}(\w)\times H(\w,r_s)\;, \label{GRads3}
\ee
we have
\ba
G^\mathrm{thermal}(\w)&=&-\frac{\Gamma^2\(-i\hw+\frac{\Delta_+}{2}\)\Gamma \(\nu\)}{\Gamma^2\(-i\hw+\frac{\Delta_-}{2}\)\Gamma \(\nu\)}\;, \nonumber\\
H(\w,r_s)&=&\frac{1+\frac{c_-}{c_+}\Big|_{r_s}\frac{\mathcal{B}_\mathrm{out}}{\mathcal{B}_\mathrm{in}}}{1+\frac{c_-}{c_+}\Big|_{r_s}
\frac{\mathcal{A}_\mathrm{out}}{\mathcal{A}_\mathrm{in}}} \; , \nonumber\\
\frac{\mathcal{B}_\mathrm{out}}{\mathcal{B}_\mathrm{in}}&=&\frac{\Gamma(1+2i\hw)\Gamma^2\(-i\hw+\frac{\Delta_-}{2}\)}{\Gamma(1-2i\hw)\Gamma^2\(i\hw+\frac{\Delta_-}{2}\)}\;, \nonumber \\
\frac{\mathcal{A}_\mathrm{out}}{\mathcal{A}_\mathrm{in}}&=&\frac{\Gamma(1+2i\hw)\Gamma^2\(-i\hw+\frac{\Delta_+}{2}\)}{\Gamma(1-2i\hw)\Gamma^2\(i\hw+\frac{\Delta_+}{2}\)}.
\ea
In fig.~\ref{Einstein}, we display the behavior of the relative deviation of the spectral density, $R(\w,r_s)$, for various values of $r_s$. A behavior qualitatively very similar to the vector case is observed, with damped oscillations in the spectral function as well as a top-down type thermalization pattern. A study of the flow of the corresponding quasinormal modes (not displayed here) is furthermore seen to fully agree with the findings of \cite{Danielsson1,Danielsson2}.

\end{appendix}

\section{Relating the Wightman function to the retarded correlator} \label{fdt}

An integral step in our evaluation of the dilepton production rate was to relate the vector field Wightman function to the corresponding retarded correlator in eq.~(\ref{fdt2}). The Wightman function can be expressed in terms of the $G_{12}$ correlator of the real time formalism through $G^{<}(\omega) = ie^{-\omega/(2T)} G_{12}(\omega)$,\footnote{Here, we have chosen the imaginary part of the lower Schwinger-Keldysh contour to be $\sigma = \beta/2$.} and a special case of the Fluctation Dissipation Theorem is then required to relate the latter to the retarded Green's function. Though this is by far not a trivial identity (for a counterexample, see ref.~\cite{CheslerTeaney}), we will argue that it holds in our quasistatic approximation, where at each stage of its motion the falling shell is treated as a stationary object residing at some radius $r_s>r_h$.

To inspect the above statements explicitly, we follow the treatment of a scalar field by Herzog and Son \cite{Herzog}, modifying it slightly to accommodate the fact that unlike in thermal equilibrium, our field exhibits both infalling and outgoing (with respect to the horizon at $r=r_h$) components outside the shell radius. To this end, we generalize eq.~(3.12) of \cite{Herzog} to read
\begin{equation}
\phi_\mathrm{outside}(\w,r)=\sum_{k} \bigg\{\alpha_k (c_+^*u_{2,k}+c^*_- u_{3,k})+
\beta_k (c_+ u_{4,k}+c_- u_{1,k}) \bigg\}\, ,
\end{equation}
where we have taken into account that the mode functions $u_{2,k}$ and $u_{3,k}$ (for definitions, see \cite{Herzog}\footnote{Note that the functions $u_{k,R,\pm}$ defined in this reference correspond to our $\phi_{\substack{\mathrm{out}\\\mathrm{in}}}$.}) are defined with negative frequency and have subsequently used our eq.~(\ref{comprel}). The computation then proceeds in perfect analogy with \cite{Herzog}, except that in their eq.~(3.16) the functions $f_k$ are to be replaced by
\begin{equation}
F_k(r)=\frac{c_+^* f_k(r)+c_-^*f_k^*(r)}{c_+^*+c_-^*} \, .
\end{equation}
In particular, eqs.~(3.17) and (3.18) of \cite{Herzog} are seen to hold, and hence we arrive at their relations (2.11)--(2.14) for the Green's functions, where the retarded correlator can be evaluated as in our eq.~(\ref{Gret}).


\end{document}